\documentclass[12pt]{article}
\usepackage{amsmath,amsfonts,amssymb,graphicx,epsfig}
\usepackage[latin1]{inputenc}
\usepackage[dvips]{color}
\def\ba{\begin{eqnarray}}
\def\ea{\end{eqnarray}}
\def\nn{\nonumber}

\begin{document}
\title{\textbf{Heisenberg model on a space with negative curvature: topological spin textures on the pseudosphere}}
%%%%%%%%%%%%%%%%%%%%%%%%%%%%%%%%%%%%%%%%%%%%%%%%%%%%%%%%%%
\author{\small{{L.R.A. Belo} $^1$\thanks{E-mail: leandrofisica@vicosa.ufv.br} , {N.M.
Oliveira-Neto}$^{1,\, 2}$\thanks{E-mails:
nmon@ufv.br, nmon@cbpf.br} , {W.A. Moura-Melo}$^1$\thanks{E-mail: winder@ufv.br} , {A.R.
Pereira}$^{1,\, 3}$\thanks{E-mail:
apereira@ufv.br} , and E. Ercolessi$^{3,4}$\thanks{E-mail: ercolessi@bo.infn.it}} \\ \\
\small
$^1$ \it Departamento de F\'{\i}sica, Universidade Federal
de Vi\c{c}osa\\ \small\it 36570-000, Vi\c{c}osa, Minas
Gerais, Brazil\\
\small $^2$ \it Centro Brasileiro de Pesquisas F\'{\i}sicas\\ \small \it Rua Xavier Sigaud, 150, Urca, 22290-180, Rio de Janeiro, Brazil\\
\small $^3$ \it Physics Department, University of Bologna\\ \small \it Via Irnerio 46, I-41126, Bologna, Italy\\
\small $^4$ INFN and CNISM\\
\small Via Irnerio 46, I-40126, Bolonga, Italy}

 \date{}
\maketitle
\begin{abstract}
Heisenberg-like spins lying on the pseudosphere (a 2-dimensional
infinite space with constant negative curvature) cannot give rise
to stable soliton solutions. Only fractional solutions can be
stabilized on this surface provided that at least a hole is
incorporated. We also address the issue of `in-plane' vortices, in
the $XY$ regime. Interestingly, the energy of a single vortex no
longer blows up as the excitation spreads to infinity. This yields
a non-confining potential between a vortex and a antivortex at
large distances so that the pair may dissociate at arbitrarily low
temperature.
\end{abstract}
\newpage
%\pacs{faltam os pacs number!!}

\section{Introduction}
As nanoscience and nanotechnology advance, materials with
astonishing small sizes have appeared. Not only their sizes, but
also their geometries have experienced a great challenge. Indeed,
besides usual ones (cylinders, cones, spheres, and so forth), the
manipulation of somewhat `exotic shapes', like the M\"obius
stripe, have been also recently reported \cite{mobius}. Actually,
the relevance of geometric and/or topological features of the
physical and/or the internal spaces has a long history in Physics.
So, non-linear topological excitations (solitons, vortices etc)
are important for understanding several properties of the system.
For example, the depairing of vortices is intimately related to
the so-called topological phase transition in a number of
quasi-planar physical systems. On the other hand, the presence of
solitons is traced back to a finite correlation length regime and
the absence of any finite temperature phase transition. Therefore,
we may ask ourselves how the underlying geometry affects the
structure, stability and dynamical properties of topological
objects and eventually some physical aspects of the system
associated to them. Indeed, a number of works has addressed such
an issue in the last years. For instance, in the context of
magnetism, several aspects of solitonic solutions associated to
the non-linear $\sigma$ model (NL$\sigma$M; which is the continuum
limit of the classical isotropic Heisenberg model) have been
studied in some geometries, like cylinders
\cite{cylinders1,Saxena-PhysA,Saxena-prb}, cones\cite{cone2}, and
so on, while a study of vortex-like excitations on a conical
background has appeared in Ref.\cite{cone-vortex}. In all of these
works, it became evident the influence of the underlying geometry
on the features of such objects. In addition, solitons have also
been investigated in non-simply connected surfaces, like the
punctured plane and the truncated cone\cite{truncated-cone,grupo}.
There, it has been verified that a fractional ($\pi/2$ or half-)
soliton emerges as the simplest non-trivial static solution of the
sine-Gordon equation, whose energy is exactly one half of that
associated with the usual $\pi$-soliton. It should be stressed
that the study of such kind of excitations, in highly non-linear
theories, is also important for biophysical and biological
processes\cite{vesicles}.

It is noteworthy that not only {\em usual geometries} have
attracted the attention of Condensed Matter physicists. As we have
already mentioned in the very beginning, samples of single
crystals with M\"obius stripes shape came to reality a few years
ago. More recently, hyperbolic spaces (with negative Gaussian
curvature) have also been considered in connection with Condensed
Matter and Statistical Physics. For instance, in Ref.
\cite{2deg-pseudosphere} the two-dimensional electron gas was
studied on the pseudosphere (the simplest hyperbolic surface,
whose curvature is constant), while in the works of
Refs.\cite{shima} the authors studied the thermodynamics of the
two-dimensional Ising model on this support, finding deviations in
some of the critical exponents associated to the negative
curvature. Here, we would like to consider a system of classical
spins described by a continuum version of the Heisenberg model
defined on the surface of a pseudosphere, in particular, how its
geometrical features affect the structure of soliton and
vortex-like excitations. Actually, we have realized that although
the pseudosphere is infinite, the isotropic model does not support
stable solitonic solutions: the negative curvature of the surface
prevents the complete mapping of the spin sphere to the physical
manifold. Therefore, stability of such excitations demands a
non-trivial topological feature, like a hole on the surface, which
avoids the collapse of the soliton. Furthermore, we also consider
vortex-like solutions in the XY regime of the former model. Now,
we have seen that the single-vortex energy is asymptotically
`regularized', i.e., its large distance term vanishes as long as
the vortex spreads to infinite. It is also verified that a
vortex-antivortex pair no longer experiences a confining potential
asymptotically, it rather appears to be free at large distances so
that entropy is expected to dominate at any arbitrary low
temperature.

\section{The model on the pseudosphere}
Let us consider the Heisenberg model for
nearest-neighbor interacting spins on a two-dimensional lattice,
given by the Hamiltonian below:
 \ba H_{\rm latt}=-J'\,\sum_{<i,j>} {\cal
H}_{i\,,j}=-J'\,\sum_{<i,j>} (S^x_iS^x_j+S^y_iS^y_j+(1+\lambda) \,
S^z_iS^z_j)\,, \label{Hlatt} \ea where $J'$ is the exchange
coupling between nearest-neighbor spins and
${\vec{S}}_i=(S^x_i,S^y_i,S^z_i)$ is the spin operator at site
$i$. The parameter $\lambda$ accounts for the anisotropy
interaction amongst spins: for $\lambda>0$ spins tend to align
along the $z$-axis (easy-axis regime); for $\lambda=0$ we have the
isotropic case, while for $-1<\lambda<0$ one gets the easy-plane
regime. Finally, $\lambda=-1$ yields the so-called $XY$ model.

In the continuum approach of spatial and spin variables, valid at
sufficiently large wavelength and low temperature, the model above
may be written as ($J\equiv{J'/2}$):

\ba
&H_1&=J\int\int\,\sum_{i,j=1}^{2}\;\sum_{a,b=1}^{3}\, g_{ij}h_{ab}\,
(1+\delta_{a3}\,\lambda)\left( \frac{\partial S^a}{\partial \eta_i} \right)\,\left( \frac{\partial S^b}{\partial \eta_j} \right)\sqrt{|g|} d\eta_1 d\eta_2 \nn\\
& & =J\int_\Omega\int\, (1+\delta_{a3}\,\lambda)(\vec{D}\,S^a)^2
d\Omega\,,\label{H1} \ea where $\Omega$ is the surface with
curvilinear coordinates $\eta_1$ and $\eta_2$ so that
$d\Omega=\sqrt{|g|}d\eta_1 d\eta_2$, $\vec{D}$ is the covariant
derivative, $\sqrt{|g|}=\sqrt{det[g]}$ and $g_{ij}$ and
$h_{ab}=\delta_{ab}$ ($\delta_{ab}$ is the usual Kronecker symbol)
are the elements of the surface and the spin space metrics,
respectively. In the static regime, expression (\ref{H1})
describes the classical properties of a number of ferro and
antiferromagnets depending on whether the sign of $J$ is positive
or negative, respectively. The Hamiltonian above is an anisotropic
non-linear $\sigma$ model (NL$\sigma$M), lying on an arbitrary
two-dimensional geometry, so that our considerations could have
some relevance to other branches like Theoretical High Energy
Physics and Cosmology.

\begin{figure}[!h]
\centering \hskip 1cm %\scalebox{0.8}
\fbox{\includegraphics[width=12cm,height=7cm]{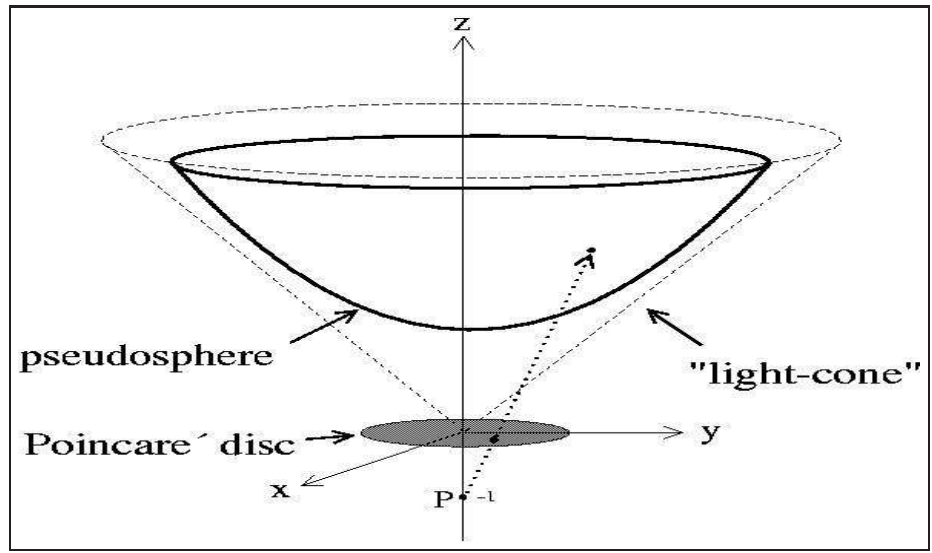}}
\caption{{\protect\small The Poincar\'e disc method to obtain the
pseudosphere (the upper hyperboloid sheet) from the projection
point $P=-1$ on the $z$ axis. Each point of the disc $(r,\varphi)$
is mapped to another on the sheet $(\tau, \varphi)$, so that the
disc border is taken to infinite.}} \label{fig1}
\end{figure}

We shall deal with such a model on the pseudosphere, which is the
simplest hyperbolic space, since it presents constant and negative
(Gaussian) curvature. Let us briefly describe some of its
geometrical features (further details may be found, for example,
in Ref.\cite{pseudosphere-review}). First of all, let us recall
that a sphere can always be embedded in a three-dimensional (3D)
Euclidian space, such that $x^2 +y^2+z^2={\cal R}^2$
(${\cal{R}}^2>0$; in Cartesian rectangular coordinates). This
global embedding cannot be generally performed for hyperbolic
surfaces. Actually, in our case the manifold is defined by
$x^2+y^2-z^2=-R^2$ ($R^2>0$), leading to two disjoint
hyperboloids. For definiteness, we choose the pseudosphere to be
the upper sheet limited by the upper ``light-cone''
$x^{2}+y^{2}=z^{2}$, $z>0$ (Figure \ref{fig1}). There are several
ways of ``visualizing" the pseudosphere\cite{pseudosphere-review}.
Here, we shall employ the Poincar\'e disc method: each point of
this disc, whose radius is $R$, is mapped to an unique point on
the pseudosphere (Figure \ref{fig1}). More explicitly, a point
parametrized by $(r,\varphi)$ on the disc is mapped to  $(x;y;z)=
(R\sinh(\tau)\cos(\varphi);\, R\sinh(\tau)\sin(\varphi);
R\cosh(\tau))$ on the pseudosphere, by means of:

\ba r=R\frac{\sinh(\tau)}{1+z/R}=R\tanh(\frac{\tau}{2})\,,\qquad
\varphi=\arctan(\frac{y}{x})\,.\label{pseudo-coord} \ea

\noindent Then $\tau\in[0,+\infty)$ and $\varphi\in[0,2\pi]$,
while the geodesic distance on the pseudosphere reads $s=R\tau/2$.
It should be noticed that both the pseudosphere and the Poincar\'e
disc present the same constant negative curvature, $-1/R^2$, as it
is expected since they are topologically equivalent. The
difference between them relies on the fact that the pseudosphere
is an infinite surface while the disc is bounded at $r=R$. In
addition, the line element of the pseudosphere reads:
\begin{equation}
ds^2 = R^2(d\tau^2+\sinh^2(\tau)d\varphi^2)\,,\label{lineelement}
\end{equation}
so that its metric tensor elements are $g_{\tau\tau}=R^2$,
$g_{\varphi\varphi}=R^2\,\sinh^2(\tau)$ and
$g_{\tau\varphi}=g_{\varphi\tau}=0$. Alternatively, we have:
$ds^2=4\left(1-\frac{r^2}{R^2}\right)^{-2}(dr^2+r^2d\varphi^2) = 4
\left(1-\frac{(x^2+y^2)}{R^2}\right)^{-2}(dx^2+dy^2)$, in usual
polar and Cartesian coordinates, respectively. Now, the
Hamiltonian (\ref{H1}) on the pseudosphere can be written as
follows:

\ba &H_2=&J\int^{2\pi}_0 d\varphi\int^\infty_0 d\tau
\left\{\sinh(\tau)\left[(1+\lambda\sin^2\theta)(\partial_\tau
\theta)^2 +\sin^2\theta(\partial_\tau \Phi)^2 \right]+\right.\nn\\
& & \hskip 3cm
\left.+\frac{\left[(1+\lambda\sin^2\theta)(\partial_\varphi
\theta)^2 +\sin^2\theta(\partial_\varphi \Phi)^2
\right]}{\sinh(\tau)} \right\}\,, \label{Hps} \ea where
$\partial_i=\partial/\partial\eta_i$. The functions
$\theta=\theta(\tau,\varphi)$ and $\Phi=\Phi(\tau,\varphi)$ are
the spin angle variables, say, $\vec{S}=(\sin\theta\cos\Phi; \,
\sin\theta\sin\Phi; \, \cos\theta)$ so that
$|\vec{S}|^2=S^2\equiv1$. From Hamiltonian above we can derive the
static Euler-Lagrange equations for $\theta$ and $\Phi$, that
read, respectively:

\ba
&\hspace{-.5cm}(1+\lambda\sin^2\theta)\left[\partial_\tau(\sinh\tau\, \partial_\tau \theta)+\frac{\partial^2_{\varphi}\theta}{\sinh\tau}\right]=&\lambda\sin\theta\,\cos\theta \left[\sinh\tau (\partial_\tau \theta)^2 +\frac{(\partial_\varphi\theta)^2}{\sinh\tau} \right]+\nn\\& & \hspace{-0.5cm}+\sin\theta\,\cos\theta \left[\sinh\tau (\partial_\tau \Phi)^2 +\frac{(\partial_\varphi\Phi)^2}{\sinh\tau} \right]\,,\label{eq1}\\ \nn\\
& \partial_\tau (\sinh\tau \, \sin^2\theta \partial_\tau\Phi)
+\partial_\varphi\left(\frac{1}{\sinh\tau} \sin^2\theta
\partial_\varphi \Phi\right)=&0\,.\label{eq2} \ea As expected,
such static equations are highly non-linear and no general
solutions for them are known up to the present. For proceeding
further in our analysis we seek for special solutions.

\section{Solitons on the pseudosphere}

In order to obtain any suitable solution for
Eqs.(\ref{eq1}-\ref{eq2}), we must impose some properties on them.
First of all, we take the isotropic case, $\lambda=0$. In
addition, let us assume that $\theta$ and $\Phi$ describe a
cylindrically symmetric solution, say, $\partial_\varphi \theta=0$
and $\Phi=\Phi(\varphi)=\varphi$ (up to a constant). With such
requirements, Eq.(\ref{eq2}) identically vanishes while
(\ref{eq1}) reduces to: \ba
\partial_\tau (\sinh\tau\, \partial_\tau \theta)=\frac{\sin(2\theta)}{2\, \sinh\tau}\,.\label{eq3}
\ea Defining $u\equiv\ln(\tanh(\tau/2))$ so that $u\in(-\infty,0]$
(this maps the pseudosphere to a semi-infinite cylinder with axial
coordinate $u$), we get a sine-Gordon equation: \ba
\partial^2_u\theta=\frac{\sin(2\theta)}{2}\,,\label{eq4sG}
\ea whose simplest solution reads\cite{Rajaraman}: \ba \theta(u)=
2\arctan(e^{{u-\overline{u}}})\,\label{sol1a} \ea where
$\overline{u}\in(-\infty,\infty)$. Its energy is easily evaluated
and gives: \ba
E_\theta=2\pi{J}\int^0_{-\infty}\left[(\partial_u\theta(u))^2+
{\sin^2\theta(u)}\right]du={8\pi{J}}\frac{e^{2\overline{u}}}{1+e^{2\overline{u}}}\,\in\,
[0,8\pi{J}]\,.\label{Esol} \ea Thus, the symmetric soliton
solution lying on the pseudosphere is unstable and always decays
to the ground-state, $E=0$. Even though the pseudosphere is an
infinite manifold, its negative curvature identifies it with the
Poincar\'e disc, which like all finite discs cannot support
Belavin-Polyakov-like excitations.\footnote{Note, however, that
the sphere does support these solitons\cite{Saxena-prb}: the spin
sphere covers the physical surface exactly $N$ times, which is
bound and compact but (stereographically) equivalent to the
compactified infinite flat plane.} Note also that if the bottom
`pseudosphere' could be connected to the upper (our actual
pseudosphere), then we would have a stable soliton. Thus, a
`major' negative curvature surface can, in principle, support this
kind of excitation (the non-disjoint `space-like' sheet defined by
$x^2+y^2-z^2=R^2>0$ is an example).

If we relax the cylindrically symmetric requirement on the
solution, say, $\partial_\varphi \theta\neq0$, then we now have
the following differential equation (with $\lambda=0$):
$\partial^2_\varphi \theta +\partial^2_u \theta=\sin(2\theta)/2$,
with $u$ defined like above (this takes the time-dependent
sine-Gordon equation if we replace $\varphi= i t$). Its simplest
solution reads: \ba \theta_{u\varphi}=2\arctan\left[
\frac{\sin(a\varphi/\sqrt{1-a^2})}{a\cosh[(u-\overline{u})/\sqrt{1-a^2}]}\right]\,,
\ea which is well-defined provided that $a/\sqrt{1-a^2}\equiv m$
is an integer. The parameter $a$ is (for $\varphi=it$) the ``speed
of the excitation" divided by the ``speed of light", so that
$a\in[0,1]$, while $\overline{u}$ is related to its radius, like
in the former case. In addition, note that: \ba
& & \theta_{u\varphi} \to 0 \quad {\mbox as} \quad u\to -\infty\,,\\
& & \theta_{u\varphi} \to
2\arctan\left[\frac{\sin(m\varphi)}{a\cosh(-ma\overline{u})}\right]
\quad{\mbox as} \quad u\to 0^-\,, \ea and, therefore, no complete
mapping of the spin sphere to the physical support is possible.
Indeed, since $\cosh(x)\ge1$ then $\theta_{u\varphi}$ never equals
$\pi$. Hence, like their cylindrically symmetric counterparts the
solution above is unstable and decays.

Actually, stability for these excitations may be provided by means
of topological obstructions. For example, if we consider a
circular-type hole with radius $s_0=R\tau_0$ centered at the
origin (equivalently, we remove a disc of radius
$r_0=R\tanh(\tau_0/2)\in(0,R)$ from the Poincar\'e disc), then
Eq.(\ref{eq3}) or (\ref{eq4sG}) is now solved, with the
requirement that $\theta(u=r_0)=\theta(\tau=\tau_0)= {\mbox \rm
constant}$, by (see, for example, Ref. \cite{truncated-cone}): \ba
\theta_{\tau_0}=2\arctan\left(\frac{\tanh(\tau/2)}{\tanh(\tau_0/2)}\right)
\,, \label{soltruncated} \ea whose associated energy reads: \ba
E|_{\tau_0}=2\pi{J}\int^\infty_{\tau_0} d\tau
\left[\sinh(\tau)\,(\partial_\tau\theta_{\tau_0})^2
+\frac{\sin^2(\theta_{\tau_0})}{\sinh(\tau)}\right]=4\pi{J}\left(1-\frac{2}{1+{\rm
cotanh}(\tau_0/2)}\right)\in (0,4\pi{J})\,. \ea In this case, we
have a fractional solution with charge interpolating between a
null ($\tau_0\to\infty$) and a $\pi/2$-soliton ($\tau_0\to0$)
whose energy runs from $0$ to $4\pi{J}$, depending on the
excitation (or on the hole) size, since scale invariance no longer
holds. Note that, the hole on the support now prevents the
collapse of the soliton, similarly to what happens in the annulus
and finite truncated cone\cite{truncated-cone}.\\

\section{Vortex-like excitations on the pseudosphere}

A magnetic vortex is commonly thought to be a spin profile with
non-vanishing vorticity. Indeed, its core may develop spin
component out of the surface whenever this is demanded for
regulating its energy. However, the XY anisotropy is frequently
invoked for ensuring that spins will lie on the surface. This
holds for the planar case, once that the $Z$-axis of the (internal
space) spin sphere is everywhere aligned perpendicular to the
surface. Therefore, spins lying on the internal equator (XY plane)
is equivalent to lie on the physical surface. Nevertheless, this
is not valid for an arbitrary surface curvature, torsion, and
other geometrical features. In these cases, a new type of
anisotropy may be demanded. For example, a term like
$b(\hat{n}\cdot\vec{S})^2$, with $b>0$ and $\hat{n}$ being a unity
vector directed normally to the surface everywhere, can guarantee
that the spins will not develop component out of the surface,
provided that $b$ is large enough (for further details, see, for
example, Ref. \cite{cone-vortex}).

A question naturally arises: how is the scenario in the present
surface? Although the pseudosphere is curved, it shares a special
property with the plane: their metric are conformal each
other\cite{pseudosphere-review} (the sphere, $S^2$, also belongs
to this group). This is to say that the difference between them
relies in measuring distance along geodesics in both supports, so
that while on the plane we have straight lines as the least way
between two distinct points, on the pseudosphere they are joined
by a hyperbole. In order to simplify our analysis, we shall employ
in this section the equivalence between the pseudosphere and the
Poincar\'e disc, which is a disc of radius $R$ endowed with a
metric that yields a constant and negative curvature. Once our
calculations are performed on this disc, where the centers of the
vortices are better identified by using Cartesian coordinates, we
move to pseudospherical coordinates for finding further properties
of the solutions on the actual support.

Taking $\lambda=-1$ in the Hamiltonian (\ref{Hps}) and its
associated differential equations, the spins will tend to align
only on the (internal) $XY$ plane and equivalently on the surface
of the Poincar\'e disc. Now, in order to look for the simplest
in-plane solution, we take $\theta=\pi/2$ and
$\partial_\tau\Phi=0$. For such a case, the differential equations
(\ref{eq1})-(\ref{eq2}) are highly simplified, leaving us
with\footnote{More precisely, we have that $\nabla^2\Phi=0$. In
pseudospherical coordinates $(\tau, \varphi)$ the Laplacian reads:
$$\nabla^2_{\tau\varphi}\Phi(\tau,\varphi)=\frac{4}{R^2}\left( \frac{1}{\sinh\tau} \,\partial_\tau(\sinh\tau\,\partial_\tau)\,+\frac{1}{\sinh^2\tau}\, \partial^2_\varphi \right)\Phi(\tau,\varphi)\,.$$
This reduces to Eq.(\ref{eqsol}) if $\partial_\tau\Phi=0$.
Equivalently:
$$
\nabla^2_{xy}=\frac14 \left(1-\frac{x^2+y^2}{R^2}\right)^2\,\left(\partial^2_x+\partial^2_y \right)\,,\qquad \nabla^2_{r\varphi}=\frac14 \left(1-\frac{r^2}{R^2}\right)^2\,\left(\frac{1}{r}\partial_r(r \partial_r)+\frac{1}{r^2}\partial^2_\varphi\right)\,,
$$
in Cartesian and polar coordinates, respectively.}

\ba
\partial^2_\varphi\Phi=0\qquad \Longrightarrow \qquad \Phi(\varphi)=Q\varphi +\varphi_0\,,\label{eqsol}
\ea

\noindent where $Q$ is the charge of the vortex (its vorticity)
centered at the origin while $\varphi_0$ is a constant related to
its global profile. The energy is easily calculated, and reads:

\ba
E_v= E_c +J\int^{2\pi}_0\,d\varphi\int^{\tau_{L}}_{\tau_{a}}\,d\tau\, \frac{1}{\sinh(\tau)}(\partial_\varphi\Phi)^2=E_c +2\pi\,JQ^2 \ln \left( \frac{\tanh(\tau_{L}/2)}{\tanh(\tau_{a}/2)} \right)\,.\label{Ev}
\ea

\noindent Here, $E_c$ is the vortex core energy which diverges in
the continuum limit for in-plane solutions (for out-of-plane
vortices, it can be estimated analytically likewise usual cases);
$\tau_{a}$ and $\tau_{L}$ are the small and large scale cutoff
parameters, related to the sizes of the inner core and outer
region of the excitation, respectively, by $a=R\tau_a/2$ and
$L=R\tau_L/2$. As it is well-known, at least in planar and conical
geometries, the vortex energy logarithmically blows up as the
vortex spreads without limit to infinite (an infrared-like
divergence). What should be stressed is that in the present case
such a divergence is naturally ruled out by means of the negative
curvature of the geometrical support. Actually, as long as
$\tau_{L}$ is raised the vortex energy increases but goes
asymptotically to:

\ba
E_v|_\infty=+2\pi\,JQ^2\ln\,({\rm cotanh}(\tau_{a}/2))\,,\label{Evinfty}
\ea
%%%%%%%%%%%%%%%%%%%%%%%%%%%%%%%%%%
\begin{figure}[!h]
\centering \hskip 1cm %\scalebox{0.8}
\fbox{\includegraphics[width=12cm,height=7cm]{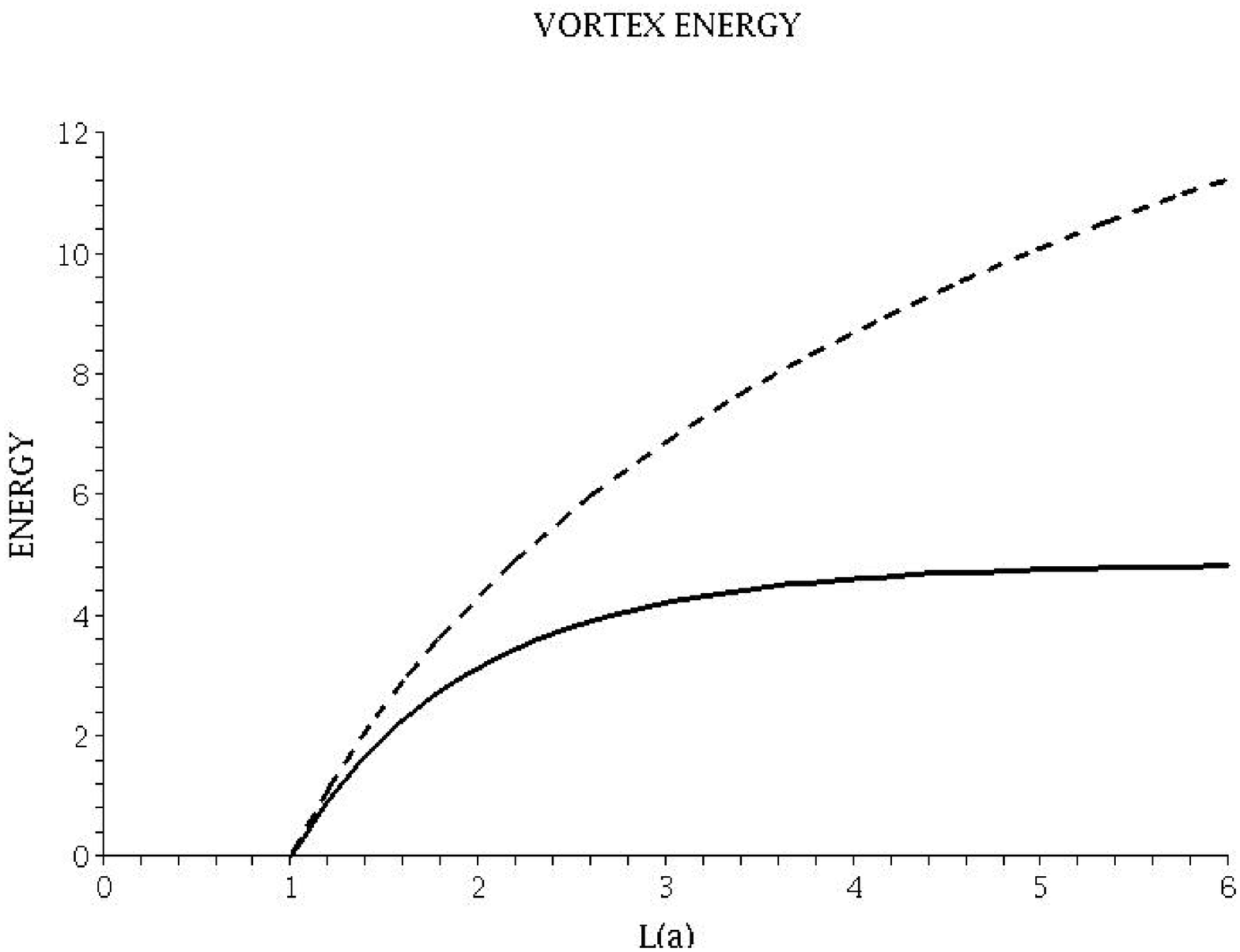}}
\caption{{\protect\small The energy of a single vortex (in units
of $JQ^2$) as a function of its outer size $L$. Upper (dashed)
curve illustrates the usual planar-like case, while the another
concerns the present geometry. Note that, in our case, the energy
goes asymptotically to $E_v|_\infty=2\pi{J}Q^2\ln2\approx4.85JQ^2$
(for $\tau_a=2a/R=1$). Therefore, its infrared-like divergence is
cured in this framework.}} \label{fig4}
\end{figure}
%%%%%%%%%%%%%%%%%%%%%%%%%%%%%%%%%%%%
which is finite (see Figure \ref{fig4}). To our knowledge this is
the first time an infrared-like divergence associated with
topological excitations is ruled out by geometrical properties.
This fact will be related to another interesting result on the
pseudosphere whenever two-vortex solutions are considered. For
that, it is more convenient to work in Cartesian $xy$ coordinates.
Since the linear combination of solutions of $\nabla^2\Phi=0$ is
another one, we take:

\ba \Phi_{2v}=Q_1\arctan\left(\frac{y-y_1}{x-x_1}\right)
+Q_2\arctan\left(\frac{y-y_2}{x-x_2}\right)\,,\label{sol2v} \ea
where $Q_1$ and $Q_2$ are the charges of the vortices centered at
$(x_1,y_1)$ and $(x_2,y_2)$ on the Poincar\'e disc (or at the
corresponding points on the pseudosphere). The energy of this
configuration may be analytically evaluated,
giving:\footnote{Instead of evaluating an integral similar to that
of Eq.(\ref{Ev}) for two vortices explicitly, we may use the fact
that $\nabla\Phi_{2v}$ is analytic everywhere, except at the
vortices centers, around which $\Phi_{2v}$ is a multivalued
function. By this method, the result of Eq.(\ref{E2v}) is much
easier and elegantly obtained. For details, see Ref. \cite{CL}.}

\ba
E_{2v}=E_{v1}+E_{v2}-2\pi{J}Q_1Q_2\ln\left(\frac{\tanh(\tau_d/2)}{\tanh(\tau_L/2)}\right)\,,
\label{E2v} \ea where $E_{v1}$ and $E_{v2}$ are the energies of
the isolated vortices (see Eq. (\ref{Ev})). The last term
represents the effective potential $V_{\rm eff}$ between the two
vortices, separated by $d=R\tau_d/2$ (measured along the
hyperbolic geodesic joining them), and $\tau_{d}\geq 2\tau_{a}$.
This potential presents a remarkable property: as
$\tau_d\to\infty$, then $V_{\rm eff}\to 0$, which is constant.
Therefore, vortices move without appreciable interaction whenever
they are sufficiently apart one from another. [This should be
contrasted to the strong logarithmic potential of usual
planar-type cases]. For the case of a vortex-antivortex pair (for
definiteness with $Q_1=-Q_2=+1$), the potential is attractive but
it plays an effective role only for enough small distances.
Therefore, even though a pair keeps tied at zero temperature, an
arbitrary weak thermal excitation may be sufficient to dissociate
it. This leads us to claim that a topological phase transition
\cite{BKT} will take
place at any temperature close to zero.\\

\section{Conclusions and Prospects}

Static and isotropic classical Heisenberg spins lying on the
geometry of a pseudosphere does not support a stable soliton.
Indeed, stability is verified if, for instance, we consider a
punctured support (a pseudosphere with a hole, rendering it a
non-simply connected feature) in which a fractional excitation
interpolating between a null and $\pi/2$-soliton is obtained.

Taking into account the XY regime, we have seen that vortices also
present an interesting property whenever compared to their usual
counterparts: their energy no longer blows up as they spread to
infinite. We have also pointed out some possible consequences of
this fact for the two-vortex solution and related issues, like
topological phase transition. [We expect that similar scenario
would take place to the so-called `out-of-plane' vortices, with
some suitable modifications for taking into account their internal
core energy]. Actually, we have seen that the cure of the infrared
divergence associated to the vortex energy on the pseudosphere
implies, for instance, that a pair of vortices no longer interacts
through a confining potential at large distances. This fact
indicates that the depairing of vortices may occur at any
arbitrary low temperature. This is another example of how the
geometry of the underlying support may affect this transition (a
previous one was provided by the conical
surface\cite{cone-vortex}). Besides of verifying this conjecture
by means of numeric/simulation techniques, other prospects for
future investigation include how solitons and vortices structure
and dynamics are affected by {\em defects}, like holes and/or
impurities, on this surface. In a more general framework it
remains the issue of how topological phase transitions are
sensitive to geometrical parameters, like curvature, torsion etc.
Some additional light to this problem could be shed by studying
the dissociation of magnetic vortices pairs on the sphere (this
study has been carried out and the results will be communicated
elsewhere\cite{workinprogress}). \vskip 1cm \centerline{\large\bf
Acknowledgements} \vskip .7cm L.A.S. M\'ol is acknowledged for
useful discussions. The authors thank CAPES, CNPq and FAPEMIG for
financial support. This work was also partially supported by the
TMR network EUCLID and the Italian MIUR through COFIN projects.

\end{document}